\documentclass[letter]{aa}  

\usepackage{graphicx,bm}
\usepackage{txfonts}
\begin{document} 

   \title{Prospects for radio detection of stellar plasma beams}
   \titlerunning{Radio detection stellar particle beams}
   \authorrunning{Vedantham}
   \author{H. K. Vedantham
          \inst{1,2}
          }

   \institute{ASTRON, Netherlands Institute for Radio Astronomy, Oude Hoogeveensedijk 4, 7991PD, Dwingeloo, The Netherlands
         \and
        Kapteyn Astronomical Institute, University of Groningen, The Netherlands 
             }

\date{Received XXX; accepted YYY}

\abstract{
Violent solar eruptions are often accompanied by relativistic beams of charged particles. In the solar context, they are referred to as SPEs (Solar Particle Events) and are known to generate a characteristic swept-frequency radio burst. Due to their ionizing potential, such beams influence atmospheric chemistry and habitability. Radio observations provide a crucial discriminant between stellar flares that do and do not generate particle beams.  Here I use solar empirical data and semi-quantitative theoretical estimates to gauge the feasibility of detecting the associated radio bursts. My principal conclusion is that a dedicated search for swept frequency radio bursts on second-timescales in existing low-frequency ($\nu\lesssim 10^2\,{\rm MHz}$) datasets, while technically challenging, will likely evidence high energy particles beams in Sun-like stars.}


   \maketitle
%
\section{Introduction}
Stellar activity can have a detrimental effect on the habitability of exoplanets. Violent releases of energy in the chromosphere and corona emit ionizing radiation and eject plasma at high speeds into the interplanetary medium. The characterise bolometric energy release in typical solar flare is about $10^{29}-10^{32}\,{\rm ergs}$ over minutes to hour-long timescales \citep{emslie-2012}. Although this is comparable to or smaller than the bolometric output of the quiescent solar disc\footnote{Hence we do not notice solar flares with our eyes.}, unlike quiescent solar emission, a large fraction of the flare energy goes into three components that are detrimental to the habitability: (a) ionizing UV and X-ray radiation, (b) ejection of a large mass of coronal plasma at high speeds (CMEs) and (c) acceleration of charges to near-light-speeds, the so-called Solar Particle Events (SPE).

Ever since the discovery of solar radio bursts, it was suspected that the radio emission is related to violent energy releases on the Sun \citep{payne,wild}. It is now well established that certain types of solar radio bursts are tell-take signatures of SPE events \citep{winter, miteva-2017}. In this paper, I concentrate on SPEs and the unique value of radio observations in evidencing their nature on stars other than the Sun. 

SPEs from solar flares usually carry about 5\% of the total flare energy \citep{emslie-2012}. The particles range from keV-level suprathermal energies to GeV-scale relativistic energies \citep{schwenn}.
Broadly speaking, the beam particles can be accelerated in two ways: (a) magnetic reconnection in the chromosphere and lower corona leading to  `impulsive' events on timescales of seconds to minutes, and (b) Fermi acceleration in a shock front\footnote{Shock formation requires bulk ejection of thermal coronal plasma at super-Alfv\'{e}nic speeds and does not happen in all flares.} leading to `gradual' events on timescales of several minutes to an hour \citep{schwenn,desai-2016,cairns-2003}. 

A beam of high energy particles moving in a dense thermal plasma leads to radio emission at the fundamental plasma frequency and its second harmonic \citep[][and references therein]{benz, zheleznyakov}. The characteristic radio bursts typically associated with impulsive and gradual particle beams are called type-III and type-II bursts respectively \citep{ameri-2019}. 
The unique diagnostic value of radio bursts of type-II and type-III lies in the fact that the ambient plasma frequency must monotonically decrease as the particle beam moves radially outward from the flare site on an unbound trajectory. Plasma emission from such beams therefore sweeps down in frequency as a function of time providing clear evidence that the flare accelerated particles will enter interplanetary space and influence the space weather around exoplanets. 

Such radio bursts from the Sun have been extensively studied both observationally and theoretically \citep{nancay,nita-2002,melrose-1980,hamish-2014}. Analogous radio bursts have not been detected from other Sun-like stars, primarily due to sensitivity  and time-on-sky limitations of previous searches. Furthermore, with few notable exceptions \citep[see for e.g.][]{bastian}, previous searches have almost exclusively focused on highly active M-dwarfs \citep{villadsen,osten2006,crosley2018a,crosley2018b,lynch} whose large-scale magnetic fields are about three orders of magnitude larger than the solar value. It is unclear if such stars emit interplanetary plasma beams or mass ejections at all (see below). In addition, the primary mechanism for bursts on these stars (and the underlying plasma instability) may be cyclotron emission due to their large magnetic field. In this case, the emission likely traces high energy plasma in magnetic traps rather than unbound plasma that streaming into interplanetary space and influence space-weather around exoplanets. Consequently, the flare dynamics and phenomenology of the radio emission is likely to be fundamentally different from those on Sun-like stars with weak magnetic fields.

I must at the outset qualify what `Sun-like' means in this context.
Stellar flares have been routinely detected in optical \citep{kepler,tess-2020} and X-ray observations \citep{pye-2015}. Such observations provide a reasonable estimate of the flare energy and duration but do not provide a concrete picture of the dynamics of the flare. For example, although dMe stars are prone to extremely energetic flares that are many orders of magnitude brighter than typical solar flares, they also have large-scale magnetic fields that are orders of magnitude stronger than the solar value. It is therefore entirely plausible that despite the copious amounts of energy releases, any plasma heated and/or accelerated in the flare is unable to force open the magnetic field lines and unbind itself from the star. Therefore I define Sun-like to mean stars of spectral class F, G and K whose magnetic field structure and plasma dynamics during flares are most likely to be similar to the Sun. It is vital to note a crucial caveat, however. Recent studies have shown that the Sun is at a critical transition point in the nature of its magnetic activity \citep{travis,sanders}. As such, extrapolating our accrued empirical knowledge on solar radio bursts to other Sun-like stars may be fraught. An explicit aim of searching for solar-type bursts on other stars is also to understand the evolution of solar activity over its lifetime by comparing it to an ensemble of stars of varying age.

If we hold the total radio energy constant, then a radio telescope is more sensitive to a shorter duration emission. It is therefore a prudent observational strategy to first go after type-III radio bursts from impulsive beams. The same techniques can be brought to bear on the more gradual Type-II events that are associated with coronal mass ejections. Thus motivated, the rest of this paper makes judicious predictions for the feasibility of detecting type-III bursts from Sun-like stars in our Galactic neighbourhood. 

\section{Sun as a prototypical star}
The simplest yield-estimate one can make is to assert that all target stars flare like the Sun. \citet{nancay} have collated flux-densities, source size, and brightness temperatures of isolated solar type-III bursts from a decade of solar observations. They present data at different channels ranging from 150 to 430\,MHz. Here I consider their statistics at 150\,MHz. The numbers of bursts that exceed a flux density threshold of $S$ is \citep[Fig. 12 from ][]{nancay}
\begin{equation}
\label{eqn:nancay}
    N(>S) = 51.63\,\left(\frac{S}{{\rm sfu}} \right)^{-0.69}\,\,{\rm day}^{-1}\,\,\,(10\,{\rm s}\,{\rm averages})
\end{equation}
where $1\,{\rm sfu} = 10^4\,{\rm Jy} = 10^{-19}\,{\rm ergs}\,{\rm cm}^{-2}$. The empirical relationship was based on a sample of bursts, the brightest of which reached about $\sim 10^{10}\,{\rm Jy}$. 
It is important to note that these statistics are based on observations with 10\,s exposures. The characteristic duration of type-III bursts at 150\,MHz is $\sim 1\,{\rm s}$ \citep{raoult-1980}. Therefore the measured flux-densities of the singular bright bursts at the optimal exposure are likely to be ten times larger.\footnote{Type-III `storms' with a train of bursts do occur, but storm-related bursts are typically faint \citep{hamish-2014}.}
If the same ensemble of bursts were viewed from a distance of $d_{\rm pc}$ parsec, then number counts will be
\begin{eqnarray}
    N(>S_{\rm mJy},d_{\rm pc})  &=& 0.17\, d_{\rm pc}^{-1.38}S_{\rm mJy}^{-0.69}\,\,{\rm day}^{-1}\,\,\,(10\,{\rm s}\,{\rm averages})\nonumber \\
    &=& 0.83\, d_{\rm pc}^{-1.38}S_{\rm mJy}^{-0.69}\,\,{\rm day}^{-1}\,\,\,(1\,{\rm s}\,{\rm averages)},\nonumber \\
\end{eqnarray}
where $S_{\rm mJy}$ is the observed flux-density in mJy units. 

It is interesting to first check if the same distributions of bursts can be detected from nearby stars using the LOFAR telescope \citep{lofar}. LOFAR telescope has a zenith System equivalent flux density of about 47\,Jy around 150\,MHz. Solar type-III bursts are known to have a duration of $\sim 1\,{\rm s}$ at 150\,MHz \citep{raoult-1980} and a spectral drift rate of $\approx 100\,{\rm MHz}\,{\rm s}^{-1}$ \citep{alvarez-1973}. For purposes of sensitivity calculations at 150\,MHz, we can simply assume the burst to be a wide-band source over an instantaneous bandwidth of 40\,MHz that is typical for LOFAR observations. The $5\sigma$ detection limit\footnote{For Gaussian statistics, $5\sigma$ corresponds to one false positive (noise spike) in $\sim 10^3\,{\rm hr}$ of observation with 1\,s exposures.} for a 40\,MHz bandwidth and 1\,s exposure is $S_{\rm lofar} \approx 25\,{\rm mJy}$ and $\approx 8\,$ mJy in a 10\,second exposure.

One can get an intuition for LOFAR's sensitivity by noting that the type-III bursts of the closest G-dwarf, Alpha-Centauti can be detected at a rate of one type-III burst every 16\,days of on-sky time. It is also noteworthy that the 25\,mJy threshold corresponds to a burst that would be $10^{9}\,{\rm Jy}$ if viewed from 1\,AU, and bursts of such brightness are detected by regular solar monitoring programs \citep{nancay,nita-2002}. In other words, the first extra-solar type-III burst is a `guaranteed' discovery with state-of-the-art radio instrumentation.\footnote{Unfortunately Alpha Centauri cannot be observed from LOFAR's latitude.} Moving further out, epsilon Eridani is a K-dwarf at a distance of $3.2\,{\rm pc}$, which must be observed on average, for about 50 days to score a detection. Here still the detectable bursts will be about $10^{10}\,{\rm Jy}$ when viewed from 1\,AU which is still within the bright-end of observed solar bursts. Regardless, the wait time is likely to be overly conservative because Epison Eridani, owing to its youth, is more active than the middle-aged Sun. 

Moving further out, our adopted value of $10^{11}\,{\rm Jy}$ for the brightest solar burst will be detectable from stars out to $d_{\rm pc}\approx 10$. There are about $100$ G-dwarfs and $300$ K-dwarfs in this volume \citep{recons}. Even if they all flared in the same way as the Sun, then the observed flare counts from a volume out to $d_{\rm pc}^{\rm max}$ based on equation \ref{eqn:nancay} is
\begin{equation}
    N(>S_{\rm mJy}) \approx 0.013\,S_{\rm mJy}^{-0.69}\left(d_{\rm pc}^{\rm max}\right)^{1.62}\,\,{\rm day}^{-1}\,(4\pi\,{\rm sr})^{-1}.
\end{equation}
For the LOFAR case, with a $25\,{\rm mJy}$ detection threshold, the all-sky rate of solar-brightness bursts $N \approx 0.06\,{\rm day}^{-1}\,(4\pi\,{\rm sr})^{-1}$.

To gauge detectability at distances exceeding $10\,{\rm pc}$, we will invariably have to extrapolate beyond the observed solar brightness range.

\section{The ensemble of flaring stars}
Solar flare energy distribution span many orders of magnitude. The brightest solar flares have a bolometric output of $\sim 10^{32}\,{\rm ergs}$. 
Observations of Sun-like stars have shown that about $\sim 1\%$ of stars display `super-flares' with a bolometric output of $10^{32}-10^{36}\,{\rm ergs}$ \citep{kepler,tess-2020}. It is reasonable to expect that the accompanying radio emissions will be significantly brighter because both the radio and optical emission are ultimately powered by the same engine. Here I adopt the simplest extrapolation in brightness one can appeal to--- the radio brightness scales in proportion to the optically inferred flare energy. 

Although such an extrapolation is admittedly simplistic, it is comforting to note the rough parity between the rates of solar radio bursts and solar flares. A consensus view from several solar flare observations is that the flare energy follows a power-law distribution with slope of $\approx -1$ (integral counts) and a normalisation that corresponds to about one bright flare of $\sim 10^{32}\,{\rm ergs}$ every $10^2\,{\rm days}$ \citep[][their Fig. 3]{schrijver}.
If we take the true flux-densities that would have been measured by \citet{nancay} in 1\,s exposures to be ten times greater than those reported from 10\,s exposures (equation \ref{eqn:nancay}), then the brightest type-III bursts have a flux density of $10^{11}\,{\rm Jy}$ and occur at a rate of $0.004\,{\rm day}^{-1}$. I, therefore, assert the following relationship to extrapolate the radio fluxes to more energetic `super-flares'.
\begin{equation}
\label{eqn:e2s}
    \frac{S}{2.5\,{\rm Jy}}\,{d_{\rm pc}^2} = \frac{E_{\rm flare}}{10^{32}\,{\rm ergs}}
\end{equation}

Armed with this relationship, we can readily predict the radio flux distribution from an ensemble of Sun-like stars. Let the volume density of the target stars be $\rho$ in units of per cubic parsecs, and their flare energies distributed according to $N(>E) = C_{32} (E/10^{32}\,{\rm ergs})^\alpha$ in units of ${\rm day}^{-1}\,{\rm star}^{-1}$. The time-rate of arrival of radio bursts brighter that some threshold from all stars within a distance $d^{\rm max}_{\rm pc}$ is then
\begin{equation}
\label{eqn:rate}
    N(>S_{\rm mJy},d^{\rm max}_{\rm pc}) = \frac{4\pi\rho C}{(3+2\alpha)}\left(\frac{S_{\rm mJy}}{2500} \right)^{\alpha}\left( d^{\rm max}_{\rm pc}\right)^{3+2\alpha}.
\end{equation}

The constants $C_{32}$ and $\alpha$ for a large population of stars of various spectral types have been well constrained by the {\em Kepler} emission, particularly for large flares in the range $10^{32}-10^{36}\,{\rm ergs}$. Here I adopt the rates published by \citet{kepler}.
Only $1.46\%$ of the targeted G-dwarfs showed detectable flaring activity. For this flaring sample, $C_{32} = 4.8$ and $\alpha=-0.96$\footnote{I estimate $C_{32}$ and $\alpha$ using the data-point at $10^{35}\,{\rm ergs}$ in their Fig. 15 and power-law indicies from their Fig. 3.}.  There are $0.0045$ G-dwarfs per cubic parsec in the solar neighbourhood \citep{recons}, which means the volumetric rate of G-dwarfs with detected super-flares is $\rho = 6.6\times 10^{-5}\,{\rm pc}^{-3}$. Substituting these in equation \ref{eqn:rate}. we get
\begin{equation}
\label{eqn:rateg}
    N(>S_{\rm mJy},d^{\rm max}_{\rm pc}) \approx 170 \,S_{\rm mJy}^{-0.96} \left(\frac{d_{\rm pc}^{\rm max}}{20}\right)^{1.08}\,\,{\rm day}^{-1}\,\,\,({\rm G\,dwarf}).
\end{equation}

Similarly, for K-dwarfs we have, $0.01$ stars per cubic parsec \citep{recons} and $2.96\%$ are known to display super-flares with $C_{32} = 0.55 $, $\alpha = -0.78$ \citep{kepler}. This gives $\rho = 3\times 10^{-4}\,{\rm pc}^{-3}$ and
\begin{equation}
\label{eqn:ratek}
N(>S_{\rm mJy},d^{\rm max}_{\rm pc}) \approx 48 \,S_{\rm mJy}^{-0.78} \left(\frac{d_{\rm pc}^{\rm max}}{20}\right)^{1.44}\,\,{\rm day}^{-1}\,\,\,({\rm K\,dwarf})
\end{equation}

The values diverge as $d^{\rm max}_{\rm pc}\rightarrow \infty$ because the flare-counts are sufficiently flat and we have not applied an upper bound on the brightness of radio emission. We address this next.

\section{Peak brightness of bursts}
Regardless of the flare energy, the peak brightness of bursts will be ultimately limited by the physics of the radiation mechanism. The mechanism of plasma emission from a particle beam is a two-stage process. Plasma density oscillations called Langmuir waves are first emitted by the beam particles. The Langmuir waves then scatter on ambient thermal ions or ion sound waves and are converted to electromagnetic waves that can escape the source (see \citet{benz, zheleznyakov,melrose-1980, kaplan-tsytovich} for further details).

Both stages have limits on the intensity of waves that can be sustained. First, the energy density of Langmuir waves is thought to be limited to about $w\sim 10^{-5}$ of the background thermal energy density $Nk_B T$, where $N$ is the thermal plasma density, $T$ is the temperature and $k_B$ is Boltzmann's constant \citep{benz}. Beyond $w\sim 10^{-5}$, the Langmuir waves begin to alter the dispersion relationship in the background plasma leading to a run-away refraction and collapse of wave packets. In the second stage, the brightness temperature to which the electromagnetic waves can grow, $T_b$, is limited to that of the Langmuir waves, $T_L$. This is because in the contrary situation, the reverse process (electromagnetic to Langmuir) would become dominant and bring about equilibrium between the two species of radiation \citep{melrose-1980, kaplan-tsytovich}. 

Consider a beam of particles with a small velocity spread $\Delta v_b$ around $v_b$. Langmuir waves around a wavenumber, $k_L = \omega_p/v_b$ and a spread $\Delta k_L /k_L = \Delta v_b/v_b$ will then be in resonance with the beam particles. Here $\omega_p$ is the angular plasma frequency and we have used the fact that Langmuir waves only exist in a small frequency range around $\omega_p$. Let the beam have an opening angle of $\theta_b$. The total wavenumber volume occupied by the Langmuir waves is $V_k = k_L^2\Delta k_L\theta_b^2 = \omega_p^3/v_b^2\times(\Delta v_b/v_b)\times\theta_b^2$. Since the energy density of Langmuir waves is $wNk_BT$, their Rayleigh-Jeans temperature is
\begin{equation}
    T_L = \frac{(2\pi)^3}{k_B}\,\frac{ wNk_BT}{V_k} = \frac{8\pi^3 NTv_b^4}{\omega_p^3\Delta v_b \theta_b^2}
\end{equation}
which at 150\,MHz, for a saturation value of $w=10^{-5}$, beam velocity of $v_b=0.3c$, and a typical solar coronal value of $T=2\times 10^6\,{\rm K}$ is 
\begin{equation}
\label{eqn:tlmax}
    T^{\rm max}_L(150\,{\rm MHz})  \sim 10^{18}\left(\frac{\Delta v_b/v_b}{0.1}\right)^{-1}\left(\frac{\theta_b}{0.1}\right)^{-2}\,\,{\rm K}
\end{equation}
Precise values of the velocity spread and the beam opening angle are not directly accessible even in the solar corona. As such, their normalisation points in equation \ref{eqn:tlmax} are educated guesses and the brightness temperature must be taken as a rough order-of-magnitude estimate. It is also important to note that the value of $T_{L}^{\rm max}$ in equation \ref{eqn:tlmax} agrees with that estimated with more detailed semi-quantitative estimates of \citet{melrose-1980}. They explicitly considered the growth rate of Langmuir waves due to the beam-instability to be balanced by diffusion of the waves in angle (and hence out of resonance with the beam) due to density inhomogeneities on scales larger than the waves themselves. I have instead absorbed that balance into the parameter $w\sim 10^{-5}$, which is also based on diffusion of Langmuir waves in angle due to density inhomogeneities. 

Having determined the peak brightness temperature of type-III radiation, we are faced with a fresh problem. Because only flux-density can be measured, we need to know the size of the emitter in order to make use of the brightness temperature. It is tempting to relate the transverse size of the emitter to the opening angle of the particle beam. This is however fraught in case of emission at the fundamental, because of the heightened level of wave refraction and scattering close to the plasma frequency \citep{melrose-1980}. Indeed, such propagation effects are to blame for the low observed polarised fraction of solar type-III bursts even though the emission mechanism itself predicts 100\% $o$-mode polarisation. A pragmatic way forward is to assign the measured median source size of $5'$ (FWHM at 1\,AU) measured for solar bursts at 150\,MHz \citep{nancay} and assume a constant solid-angle broadening factor of $10^2$ as inferred for solar bursts \citep{melrose-1980}.

The peak flux density of a source operating close to saturation at a distance of $d_{\rm pc}$ parsec, is then readily found to be\footnote{One can check that the adopted peak flux-density for solar bursts of $10^{11}\,{\rm Jy}$ gives $T_b \lesssim 10^{15}\,{\rm K}$ which is comfortably within the theoretical maximum for scatter broadening of the solid angle by factors up to $10^3$. }
\begin{equation}
 \label{eqn:peak_rlum}
    S_{\rm max}(150\,{\rm MHz}) \sim 35 \, \left(\frac{d_{\rm pc}}{100}\right)^{-2}\,\,{\rm mJy}
\end{equation}

In other words, given the sensitivity of existing telescopes like LOFAR at 150\,MHz, the distance-horizon for super-flares is $\sim 100\,{\rm pc}$.

We can now also bound the flux-density in our simple model of equation \ref{eqn:e2s} to relate super-flare energy to radio-burst flux density. The peak radio spectral luminosity at saturation (from equation \ref{eqn:peak_rlum}) is $\sim 350\,{\rm Jy}\,{\rm pc}^2$, or about $10^{16.5}\,{\rm erg}\,{\rm s}^{-1}\,{\rm Hz}^{-1}$. Putting this in equation \ref{eqn:e2s}, I find that the radio luminosity from metre-wave radio bursts saturates at a bolometric flare energy of $\sim 10^{34}\,{\rm erg}$. Saturation, therefore, leads us to modify equation \ref{eqn:e2s}:
\begin{eqnarray}
    S_{\rm Jy}\,d_{\rm pc}^2 &=& 2.5\,\frac{E_{\rm flare}}{10^{32}\,{\rm erg}};\,\,E<10^{34.15}\,{\rm erg}\nonumber \\
    & = & 350;\,\,E\geq 10^{34.15}\,{\rm erg}
\end{eqnarray}

Equations \ref{eqn:rateg} and \ref{eqn:ratek} remain valid so long as $d_{\rm pc}^{\rm max}$ and $S_{\rm mJy}$ are chosen to obey \ref{eqn:peak_rlum}. In summary, the emission mechanism combined with empirical constraints on apparent source size show that, to account for super-flare related emission, it is reasonable to scale the radio flux-densities up to a factor of $\sim 10^2$ from the brightest solar bursts ever observed.

\begin{figure}
    \centering
    \includegraphics[width=\linewidth]{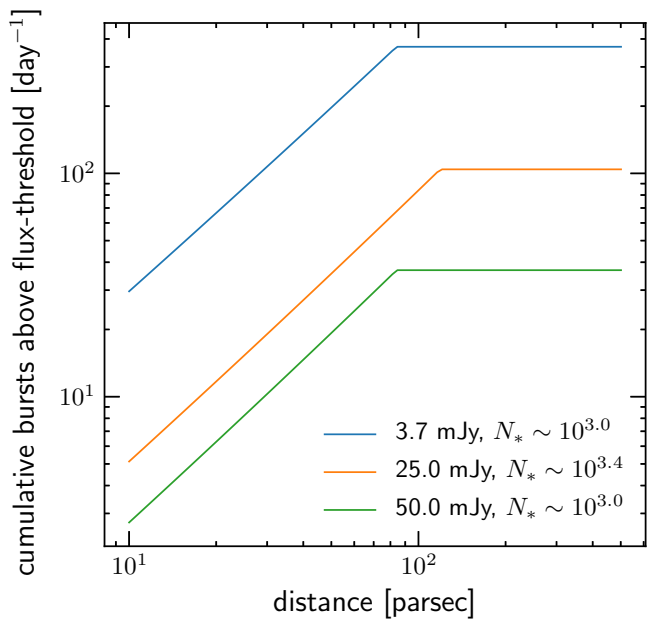}
    \caption{Expected all-sky rate of radio bursts at 150\,MHz accompanying super-flares ($E\gg 10^{32}\,{\rm erg}$) from stars within a distance given by the $x-$axis values. The different curves correspond to the flux-sensitivity thresholds given in the legend. The thresholds correspond to the planned SKA-Low sensitivity (blue), and current LOFAR sensitivity (optimistic in orange and conservative in green) in a 1\,s integration and 40\,MHz bandwidth respectively. The break in the curves is at the sensitivity `horizon' and is due to saturation of the emission process. The expected number of stars within the horizon contributing to the super-flares is also given  in the legend.}
    \label{fig:my_label}
\end{figure}
\section{Outlook}
The strategy for type-III radio bursts discovery beyond the solar system will be two-fold. (a) As more nearby ($d_{\rm pc}\sim 10$) stars' flare-statistics becomes available with TESS data, one can use the estimates presented here to identify promising Sun-like stars for targeted campaigns to go after $\sim 10^{32-33}\,{\rm erg}$ flares. These are `normal' flares for which empirical solar radio data can be applied in a comparative study. (b) The ensemble flare statistics for super-flares ($E\gg 10^{32}\,{\rm ergs}$) are already constrained by Kepler data. The larger distance-horizons for super-flares allows searches in existing archival data. For e.g. the ongoing LoTSS survey with LOFAR \citep{lotss} is surveying the northern sky at 150\,MHz using 3168 pointings, each lasting 8\,hours.\footnote{The survey data are archived at the necessary 1\,s resolution.} The instantaneous FWHM field of view of LOFAR is about $16\,{\rm deg}^2$. This amounts to $\sim10^{5.5}\,{\rm deg}^2\,{\rm hr}$ of sky exposure. Assuming a $10\sigma$ detection threshold of 50\,{\rm mJy}, we get a distance-horizon of 85\,pc, which yields an all-sky rate from equations \ref{eqn:rateg} \& \ref{eqn:ratek} of $\approx 10^{1.6}\,{\rm day}^{-1}\,(4\pi\,{\rm sr})^{-1}$ (see also Fig. 1). The bursts will originate from $N_\ast \approx 10^3\,$stars capable of super-flares, which means each LOFAR pointing will have about $0.4$ such stars on average. The total yield over the planned 3168 $8$-hr survey pointings is therefore $\gtrsim 10^1$ events which is very promising but not absolutely guaranteed due to the inherent uncertainties involved.

I end by noting two caveats. (a) The super-flare related estimates here must be taken as being order of magnitude values as they are based on semi-quantitative arguments. Although inherent uncertainty will remain until real detections are brought to bear, modelling of expected burst spectra for different coronal parameters is a fruitful avenue for future work. Solar type-III burst attain their peak brightness temperature at decametre wavelengths that are emitted at $\sim 10\,R_\odot$. For instance, the brightest type-III bursts are observed to have a flux-density of  $10^{12}\,{\rm Jy}$ at $1\,{\rm MHz}$ \citep{dulk2000}, which is an order of magnitude larger than the peak 150\,MHz flux-density adopted here. The base coronal density of stars prone to super-flares are likely much higher and although it is tempting to postulate that their peak brightness will be attained at metre-wavelengths, further theoretical work is necessary to make an informed statement.
(b) Having achieved a detection, one is still faced with the task of demonstrating the swept frequency nature of the bursts--- a necessary step to confirm the propagation of a plasma beam along an unbound trajectory. This may prove challenging in some cases where the observation bandwidth and time-resolution do not allow for the frequency-sweep to be unambiguously detected. Based on the solar experience, in such cases, type-III bursts and the so-called spike bursts may be difficult to tell apart. Spike burst are thought to be driven by a different emission mechanism--- the electron cyclotron maser which can also attain very high brightness temperatures \citep{cliver}. They however occur on much shorter timescale (millisecond to tends of millisecond) and can therefore be readily distinguished from type-III burst in data with sufficient resolution.

%

\bibliographystyle{aa}
\bibliography{ref}

\begin{acknowledgements}
I thank Prof. Gregg Hallinan for introducing me to the subject of extra-solar space weather, and Dr. Hamish Reid for a short albeit interesting discussion on this topic. I thank Dr. Ben Pope and Dr. Joe Callingham for commenting on the manuscript and the anonymous referee for helpful comments.
\end{acknowledgements}
\end{document}